\documentclass[11pt]{article}

\usepackage[english]{babel}
\usepackage[letterpaper,top=1in,bottom=1in,left=1in,right=1in,marginparwidth=0.75in]{geometry}

\usepackage[version=4]{mhchem} 
\usepackage{siunitx}
\usepackage{graphicx}
\usepackage{dcolumn}
\usepackage{bm}
\usepackage{setspace}
\usepackage{amsmath}
\usepackage{fixmath}
\usepackage{graphicx}
\usepackage{amssymb}
\usepackage{mathtools}  
\usepackage{cite}
\usepackage{float}
\usepackage{authblk}

\title{Dark-Field X-Ray Microscopy with Structured Illumination for Three-Dimensional Imaging}

\author[1]{Do\u ga G\" ursoy\thanks{dgursoy@anl.gov}}
\author[2,3,4]{Kaan Alp Yay}
\author[1,5]{Elliot Kisiel}
\author[1]{Michael Wojcik}
\author[1]{Dina Sheyfer}
\author[6]{Arndt Last}
\author[1]{Matthew Highland}
\author[3,4,7]{Ian Randal Fisher}
\author[8]{Stephan Hruszkewycz}
\author[1]{Zahir Islam\thanks{zahir@anl.gov}}

\affil[1]{Advanced Photon Source, Argonne National Laboratory}
\affil[2]{Department of Physics, Stanford University}
\affil[3]{Geballe Laboratory for Advanced Materials, Stanford University}
\affil[4]{Stanford Institute for Materials and Energy Sciences, SLAC National Accelerator Lab.}
\affil[5]{Department of Physics, University of California, San Diego}
\affil[6]{Institute of Microstructure Technology, Karlsruhe Institute of Technology}
\affil[7]{Department of Applied Physics, Stanford University}
\affil[8]{Materials Science and Engineering Division, Argonne National Laboratory}

\begin{document}
\doublespacing
\maketitle

\begin{abstract}
We introduce a structured illumination technique for dark-field x-ray microscopy optimized for three-dimensional imaging of ordered materials at sub-micrometer length scales. Our method utilizes a coded aperture to spatially modulate the incident x-ray beam on the sample, enabling the reconstruction of the sample’s 3D structure from images captured at various aperture positions. Unlike common volumetric imaging techniques such as tomography, our approach casts a scanning x-ray silhouette of a coded aperture for depth resolution along the axis of diffraction, eliminating any need for sample rotation or rastering, leading to a highly stable imaging modality. This modification can provide robustness against geometric uncertainties during data acquisition, particularly at higher resolutions where geometric uncertainties approach and can limit the target resolution. We introduce the image reconstruction model and validate our results with experimental data on an isolated twin domain within a bulk single crystal of an iron pnictide obtained using a dark-field x-ray microscope. This timely advancement aligns with the enhanced brightness upgrade of the world’s synchrotron radiation facilities, opening unprecedented opportunities in imaging.
\end{abstract}

\section*{Introduction}

Dark-field x-ray microscopy (DFXM) \cite{simons2015dark} is a developing imaging technique for characterizing the micro- and meso-scale structural, electronic, and magnetic features of ordered materials in the form of single crystals, polycrystalline materials, and epitaxial thin films, with potential applications in artificial multilayers and quasicrystals. Unlike traditional transmission geometries, which rely on grating interferometry \cite{pfeiffer2008hard} or speckle scanning \cite{wang2015hard} to obtain dark-field or scattering signals, DFXM operates in Bragg geometry and utilizes a lens-based approach. This method employs diffracted x-ray Bragg peaks to generate real-space images that highlight structural heterogeneities visible through Bragg diffraction contrast \cite{jakobsen2019mapping}, differentiating it from traditional in-line diffraction contrast techniques \cite{davis1995phase}. As a result, DFXM excels in isolating features such as structural twins, subtle charge and magnetic domain boundaries, interfaces between phases, small grains, and extended defects (e.g. dislocations) in bulk matter, which are often impractical to image for other imaging modalities such as coherent diffraction imaging, electron microscopy or scanning tip methods \cite{yildirim2020probing}.

Typically, DFXM employs an x-ray beam focused to sub-\SI{100}{\micro\meter} to illuminate a specific region in a sample oriented to excite a diffraction condition from a local grain, or domain, of interest. The Bragg diffracted beam passes through an x-ray objective and subsequently a magnified real-space intensity image is captured on a pixelated area detector, enabling diffraction-contrast images of the sample to be measured. The contrast stems from a plethora of structural factors, including variations in lattice orientation, scattering amplitude and phase, as well as lattice strain, typifying the heterogeneities existing at various length scales in ordered materials. While the microscope with this arrangement can provide diffraction contrast in projected 2D images of the sample at a Bragg condition, transitioning from a 2D to 3D image generally requires rotating the sample around a fixed momentum-transfer vector (Q) and solving the tomographic reconstruction problem to obtain a 3D image of the sample.

However, achieving accurate tomographic reconstruction requires extremely precise alignment of the sample around the Q vector to maintain the Bragg condition throughout the rotation. This demands rotation stages with minimal angular and translational runout, such as air-bearing stages, and a tedious manual alignment process. The reconstruction is further complicated by variations in the microscope's effective resolution function depending on the tomographic rotation angle \cite{poulsen2017x}. Additionally, the effectiveness of geometry alignment algorithms for the lenses \cite{breckling2022automated}, or the sample \cite{gursoy2017rapid} can be compromised if the incident beam intensity used for normalization is insufficient or fluctuates, or if the sample features change significantly at different incident angles near the Bragg peak. Due to these challenges, successful 3D reconstructions based on tomographic data remain scarce, with only one reported case to date \cite{simons2015dark}.

An alternative to rotation-based methods is raster scanning with a thin beam (less than \SI{1}{\micro\meter}) to map the sample volume. This technique relies on raster scanning of the sample, where the resolution depends on the focused beam size on the sample and the raster step. While this approach is computationally straightforward by eliminating the need for a full tomographic reconstruction, it is constrained by long collection times and potential instabilities associated with either rastering the sample or the focusing lens. Such instabilities can induce angular fluctuations that disrupt the Bragg condition, leading to significant degradation of spatial resolution in the scattering direction, particularly pronounced at small scattering angles. A method that employs a full beam without requiring sample movement could greatly enhance 3D DFXM by improving stability, reducing alignment time, and potentially enabling faster data collection.

In this paper, we propose leveraging a structured x-ray incident beam, as discussed in the case of optical light by Forbes et al. \cite{forbes2021structured}, as a robust and time-efficient alternative to conventional rotation-based DFXM for achieving 3D resolution. Structured light, widely utilized in diverse fields such as ghost imaging \cite{shapiro2008computational} and single-pixel cameras \cite{duarte2008single}, involves projecting predetermined light patterns onto the sample and capturing variations in the reflected or transmitted light using a detector. Additionally, we investigate advancements in manufacturing x-ray modulators in the form of absorbing coded apertures at micron scales and evaluate their potential applicability in realistic length scales for DFXM. This structured illumination method maintains the sample in a fixed position, circumventing the instabilities associated with sample rotation and calibration to which we alluded above, imposing fewer restrictions on sample environments, making it well-suited for conducting \textit{in situ} or \textit{operando} imaging experiments at relevant resolutions with upgraded synchrotron sources. Additionally, it can facilitate imaging of plate-like samples, such as those in our study, which present challenges due to increased thickness at certain angles, resulting in artifacts similar to those seen in limited-angle tomography. However, it is important to note that this method is susceptible to intensity fluctuations. Therefore, a stable beam or a method for monitoring beam intensity during scanning is required. To demonstrate the effectiveness of our approach, we apply it to the structural 3D imaging of the low-temperature orthorhombic phase of a single-crystal \ce{Ba(Fe_{0.98}Cu_{0.02})_2As2} sample.

\section*{Results}

\subsection*{Working Principle}

\begin{figure}
\centering
\includegraphics{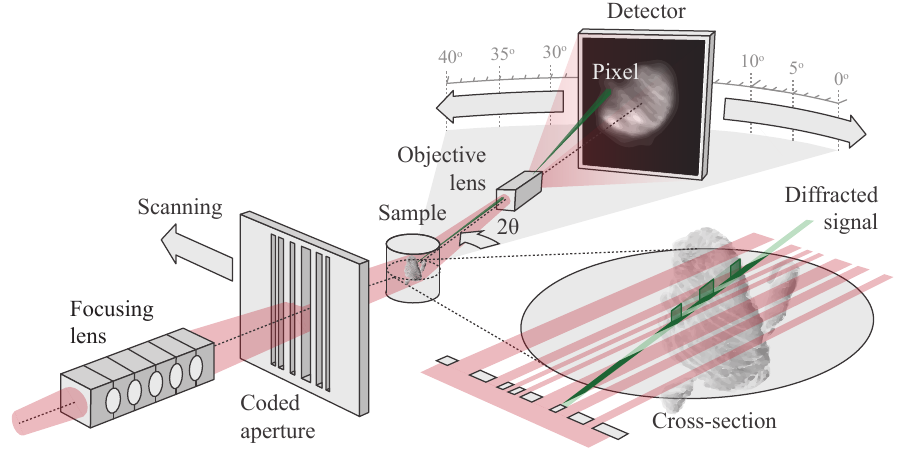}
\caption{\textbf{Experimental setup for 3D DFXM with structured illumination.} A focusing lens directs the x-ray beam onto a selected crystalline grain or domain, while a 2D pixel-array detector captures its magnified image at the Bragg angle ($2\theta$) via an objective lens. A scanning coded aperture with an absorbing micro-pattern modulates the illumination for data collection from a fixed sample, enabling digital reconstruction of the grain's 3D structure. The cross-section of the sample is magnified to illustrate how a diffracted signal in a detector pixel responds to the internal diffraction (green boxcars) of a specific grain dictated by the selectively illuminated regions.}
\label{fig1}
\end{figure}

We illustrate a representative experimental setup demonstrating the application of micro-coded apertures for structuring illumination in DFXM in Fig.~\ref{fig1}. A focused beam of illumination, ranging from about 10 to \SI{100}{\micro\meter} in diameter, is directed onto a targeted volume within a crystalline sample. The goal is to achieve an x-ray beam sufficient in size so as to illuminate the entire volume of interest without scanning. In this context, a variety of x-ray optics options are available, ranging from capillaries to refractive lenses and mirrors. The real-space image of the illuminated volume projected to 2D undergoes magnification through an objective lens and is subsequently recorded by a pixel-array area detector positioned at a required imaging distance along and normal to a Bragg-diffracted beam. Selecting an objective lens with a numerical aperture typically around 0.0004 is crucial to reject other diffracted signals. CRL \cite{snigirev1996compound} or Fresnel zone plates \cite{gorelick2011high} are excellent choices that offer this capability. 

This setup captures a 2D image of the diffracting portions of the illuminated volume onto the detector. To achieve spatial resolution along the Bragg diffracted beam, we propose structuring the focused incident beam transversely to the incident-beam propagation, and recording the modulation of intensity measurements in each detector pixel to a controlled change of the imposed structure in illumination. The aim is to encode depth information along the diffracted beam by modulating intensity and subsequently employ digital reconstruction techniques to decipher or decode the recorded intensities for each pixel.

However, structuring light in the hard x-ray regime poses significant challenges due to the fact that hard x-rays interact weakly with matter compared to visible or near-infrared light. As a result, our approach utilizes an x-ray-absorbing barcode-like structure with pseudo-random patterning as the aperture. This aperture is scanned normal to the beam to introduce diversity into the illumination, enabling depth (axial) resolution along the exit wave. Nanofabrication methods are available to produce micro-coded apertures of highly absorbing gold or tungsten structures ranging from 1 to 10 micrometers in thickness on a thin membrane. As we show, these micro-coded apertures can achieve the desired depth resolution of a few micrometers or less, approaching to the sub-micrometer transverse resolution in the image plane in our present study.

The cross-section of the sample depicted in Fig.~\ref{fig1} offers an in-depth exploration of the model of DFXM with structured illumination. Here, the coded aperture imparts a distinct pattern to the x-ray beam before its interaction with the crystalline sample. As a result, the response in a single pixel of a detector is set by diffraction occurring only at the locations marked in dark green. Essentially, the measurement at a detector pixel can be described linearly by the summation of all of those diffracting elements selected by the structured illumination. By systematically changing the coded aperture's position in known steps, we can generate a set of image measurements, each resulting from different patterns of illumination contributing to each pixel. This concept allows us to establish a system of linear equations that facilitates the determination of the diffraction profile within the grain along the direction normal to the DFXM image plane. This reconstruction process can be carried out independently for each detector pixel, thus providing a three dimensional image with a diffraction contrast mechanism dictated by the Bragg condition of the measurement.

In mathematical terms, the structured illumination can be represented by the vector $\mathbold{a} = [a_1, a_2, \hdots, a_L]^T$, with each coefficient describing the optical transmissivity of a given location of the coded aperture. Here, $L$ is the number of components in a piece-wise constant representation of the entire aperture. For example, a value of `0' indicates that an incident beam on the coded aperture can pass through without being absorbed, while `1' signifies complete absorption. In practice, these values often fall between 0 and 1, depending on factors like the thickness and material of the coded aperture, as well as the energy and incidence angle of the x-ray beam. 

Because the footprint of the x-ray beam on the aperture surface is smaller than the total aperture size, we use a smaller segment of the coded aperture, denoted as $\mathbold{a}_p = [a_{p}, \hdots, a_{p+N-1}]^T$, to model the interaction of the beam with the aperture. This can be mathematically expressed through matrix multiplication as follows:
\begin{equation}
\label{eq1}
\begin{bmatrix}
    d_{1} \\
    d_{2} \\
    \vdots \\
    d_{M}
\end{bmatrix}
=
\begin{bmatrix}
    a_{p} & a_{p+1} & \dots  & a_{p+N-1} \\
    a_{p+1} & a_{p+2} & \dots  & a_{p+N} \\
    \vdots & \vdots & \ddots & \vdots \\
    a_{p+M-1} & a_{p+M} & \dots  & a_{p+M+N-2}
\end{bmatrix}
\begin{bmatrix}
    s_{1} \\
    s_{2} \\
    \vdots \\
    s_{N}
\end{bmatrix},
\end{equation}
where $\mathbold{d} = [d_1, d_2, \hdots, d_M]^T$ represents the intensity vector acquired from a single detector pixel, which we obtain by translating the coded mask $M$ times, and $\mathbold{s} = [s_1, s_2, \hdots, s_N]^T$ denotes the signal scattered from the sample at the Bragg angle and collected in this pixel. Note that $N$ is an arbitrary number that we select and is often smaller than $L$ to cover the segment of the aperture that is used in data acquisition. We can use matrix notation to express Eq.~\ref{eq1} more concisely:
\begin{equation}
\label{eq2}
    \mathbold{d} = \mathbold{A} \mathbold{s},
\end{equation}
where $\mathbold{A}$ is the $M \times N$ coding matrix. Each row corresponds to a detector pixel at a scan point, representing the coded aperture's geometry at that moment. The complete matrix $\mathbold{A}$ encompasses the entire scanning process. 

Accurate knowledge of $\mathbold{A}$ is essential for success and must be achieved through characterization of the structure before the experiment, along with its positioning during the scan. Various microscopy techniques, such as optical, electron, or x-ray microscopy, can be employed depending on the desired resolution. Once the aperture structure is known, its spatial positioning can be coarsely determined using DFXM in bright-field mode. Finer positioning is less critical when the aperture is normal to the incident beam, allowing minor misalignments to be disregarded at the targeted resolution in this paper. For higher resolutions, more precise calibration and tracking methods such as laser interferometry can be utilized. Additionally, encoder readings from the motorized scan help in tracking aperture positions during the experiment, and this information is directly incorporated into constructing the $\mathbold{A}$ matrix.

The corresponding image reconstruction problem can be formulated in the least-squares sense. Given the measured dark-field intensity data vector, denoted as \(\mathbold{d}\), and the system matrix, denoted as \(\mathbold{A}\), representing the linear relationship between the unknown signal vector, denoted as \(\mathbold{s}\), and the measured data, the problem can be written as:
\begin{equation}
\label{eq3}
\min_{\mathbold{s}} \| \mathbold{d} - \mathbold{As} \|^2_2\
\end{equation}
where \(\| \cdot \|^2_2\) denotes the Euclidean norm. The system matrix $(\mathbold{A})$ holds the relationship between the coded aperture pattern and the resulting scattered x-ray signals. The resolution of the reconstructed signal is closely related to the rank of the system matrix and how we solve the problem in Eq.~\ref{eq3}. By solving this problem for each pixel in the detector, we can recover an image representation of the sample in the least-squares sense. 

\subsection*{Validation on a Twinned Orthorhombic Structure}

We validated our technique by imaging the 3D low-temperature microstructure of a single crystal of \ce{Ba(Fe_{0.98}Cu_{0.02})_2As2}, commonly referred to as Cu-Ba122. Cu-Ba122 belongs to the 122 family of iron pnictides known to host many interesting electronic and magnetic properties such as electronic nematicity and unconventional superconductivity \cite{paglione2010high, fernandes2022iron}. At room temperature, Cu-Ba122 possesses the tetragonal crystal structure of \ce{ThCr2Si2}\cite{rotter2008superconductivity} and upon cooling to $T_S=94\pm0.45$\SI{}{\kelvin}, it undergoes a structural phase transition from tetragonal to orthorhombic symmetry.  This transition induces an orthorhombic distortion that causes atoms to shift along the tetragonal $(110)_T$ direction. Consequently, the primitive lattice vectors along the $ab$-plane in the orthorhombic phase experience a \SI{45}{\degree} rotation compared to those in the tetragonal phase, and their lengths become unequal ($a_O > b_O$). Additionally, the formation of orthorhombic twin domains below the phase transition leads to the splitting of the tetragonal $(220)_T$ peak into four peaks, each of which corresponds to a domain type, mirroring observations in the parent compound \ce{BaFe2As2} \cite{tanatar2009direct, blomberg2012effect}.

\begin{figure}
\centering
\includegraphics{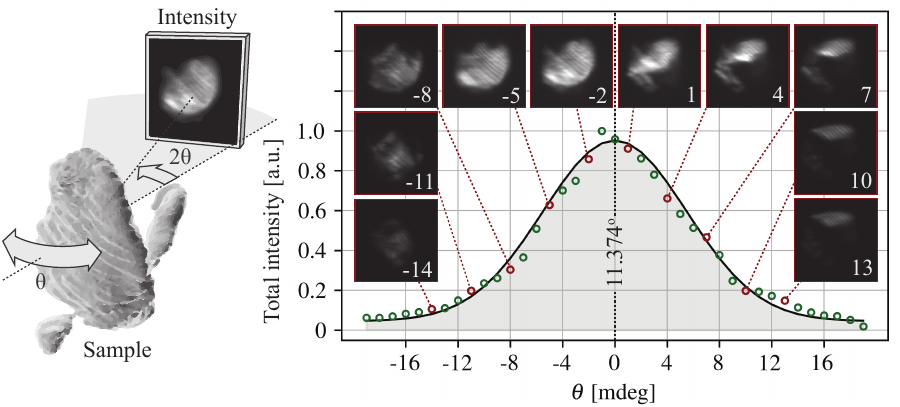}
\caption{\textbf{Recorded diffraction images at different $\theta$ angles.} Measured intensity plot versus the angle of incidence $\theta$ is obtained by integrating the images collected at the corresponding angle. The images captured without the coded aperture are recorded by incrementally stepping $\theta$ in increments of \SI{0.003}{\degree}, with the detector positioned at the Bragg angle $(2\theta)$. Fitting a Gaussian function to the Bragg peak reveals a mean of \SI{11.374}{\degree} and a standard deviation of \SI{0.0055}{\degree}.}
\label{fig2}
\end{figure}

We collected DFXM measurements at the 6-ID-C beamline of the Advanced Photon Source (APS) to capture the spatial structure of the domains resulting from this phase transition. During the measurements, we kept the sample below $T_S$ at $60\pm 0.05$K in a low-vibration cryostat \cite{plumb2023dark} and imaged the Bragg peak corresponding to one of the four orthorhombic domain types in a Laue horizontal diffraction geometry. We show the resulting DFXM images of a particular orthorhombic domain deeply embedded within the bulk crystal in Fig.~\ref{fig2}. This initial data consists of a series of 2D DFXM images taken without a coded aperture. We fixed $2\theta$ and mapped out the diffracting features of the twin domain in fine $\theta$ steps about the Bragg condition. A noteworthy feature of this particular domain is the presence of micron-scale spatial modulations in diffraction intensity. These modulations permeate throughout the image, and the axis of modulation is inclined by \SI{45}{\degree} with respect to the horizontal diffraction plane. Since the observed images are projections of the orthorhombic domain onto a 2D detector, the observed spatial contrast is indicative of the full 3D structure of the domain even if the images themselves lack depth resolution.

\begin{figure}[t]
\centering
\includegraphics{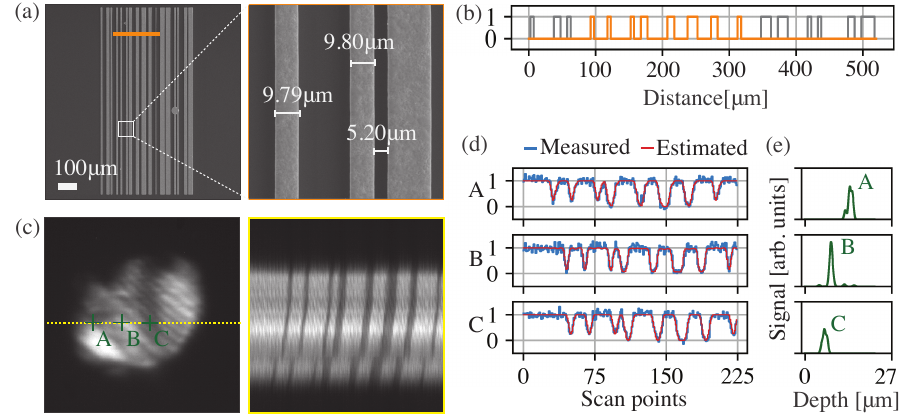}
\caption{\textbf{Depth estimation by scanning the coded aperture.} (a) SEM image of the coded aperture, with a highlighted zoomed-in section for a detailed view of the bar lengths. (b) Binary code sequence. The orange line in (a) and (b) indicates the bars utilized for encoding in our acquisition. (c) A representative single dark-field image recording on the left and a scanning image corresponding to the transverse plane on the right as marked by the dotted yellow line. The vertical axis of the right image is the direction along the dotted line, whereas the horizontal axis represents different scan steps of the coded aperture. The vertical dark region seen above the reference point C is an example of the shadows cast by the coded aperture during the scan. These shadows are removed as a result of the reconstruction algorithm. The \SI{45}{\degree}--inclined modulation in the detector image are features inherent to the sample. The shadow of an aperture bar is visible in the middle of the left image, and intensity modulation is visible in the transverse plane image on the right. Reference points A, B, and C are marked for use in the subsequent plot (d), which illustrates their measurements obtained during the scanning of the coded aperture. Corresponding estimated values derived from depth-resolved signals are presented in (e).}
\label{fig3}
\end{figure}

To achieve axial depth resolution, the collection of 3D data involved fixing $2\theta$, positioning $\theta$, and scanning the coded aperture across the incident beam at a location \SI{9.4}{\milli\meter}. Since resolution depends on the projected bit sizes at the sample's position rather than on distance alone, one could design a tailored aperture with bit sizes suited to the specific experimental configuration and constraints. Ranges of $2\theta$, $\theta$, and aperture are detailed in the Data Collection section. We note that the angles ($\theta$, $2\theta$) selected for coded aperture scans were more coarsely spaced compared to the 2D DFXM scan (see dark red marks in Fig.~\ref{fig2}). For the coded aperture, we employed a binary barcode design composed of gold (\ce{Au}) lines as shown in Fig.~\ref{fig3}a,b. Scanning this coded aperture across the incident beam cast shadows of the \ce{Au} lines on the sample and thus modified the scattering intensity captured by the detector pixels. Such a vertical shadow can be observed in the middle of the left image in Fig.~\ref{fig3}c. 

Furthermore, the full shadow profile of the coded aperture can be viewed by plotting the intensity of the pixels along the yellow dotted line in Fig.~\ref{fig3}c as a function of aperture scan position. The resulting plot shown on the right in Fig.~\ref{fig3}c displays slanted bar profiles that are indicative of the difference in the depth within the domain from which the corresponding scattering intensity emanates from, providing visual evidence for the depth sensitivity allowed by the coded aperture method. Moreover, Fig.~\ref{fig3}d depicts for three representative pixels the excellent agreement between the measured intensity and the estimated intensity as a function of scan position. The estimated intensity is derived from the depth-resolved signals resulting from the reconstruction process discussed above, and the agreement provides further quantitative corroboration of the depth sensitivity. 
 
\begin{figure*}[!t]
\centering
\includegraphics{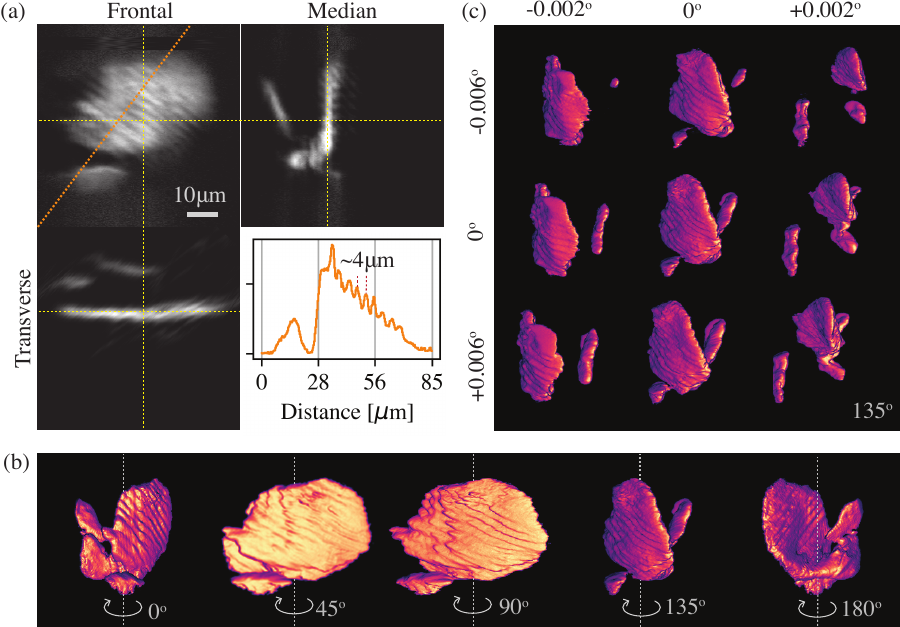}
\caption{\textbf{Reconstructed 3D volume of the imaged domain in the Cu-Ba122 sample.} (a) Orthogonal planar views depicting the reconstructed 3D structure. The plot represents a profile along the orange-marked line in the frontal view. (b) Rendered perspectives of the 3D structure from various rotational angles. (c) Renderings illustrating the reconstructed 3D structures at a select of scattering ($\Delta(2\theta)=\mp\SI{0.002}{\degree}$) and rocking angles ($\Delta\theta=\mp\SI{0.006}{\degree}$). The middle image corresponds to the one displayed in (b).}
\label{fig4}
\end{figure*}

We present our results that capture the full 3D structure of the embedded orthorhombic domain in our Cu-Ba122 sample in Fig.~\ref{fig4}. At several different near-Bragg diffraction conditions we imaged this domain in real space, here represented either as diffraction density cuts through the reconstruction (a) or as isosurfaces (b, c). These 3D reconstructed images enable us to observe that the 45$^\circ$-inclined intensity modulations in the 2D images in Fig.~\ref{fig2} permeate through the depth of the domain. Similar modulations were seen in surface-sensitive optical measurements before in related compounds, and corresponding diffraction studies suggested that the modulations extend along the depth of the sample \cite{tanatar2009direct}. We expect these modulations to arise from local deviations from the average structure of the domain, and the related mechanism will be studied in future work. Here, we emphasize that our reconstructed images are in line with previous observations and also extend our understanding by elucidating the full 3D structure of the domain.

Further to note is the excellent angular resolution afforded by our low numerical aperture that allows us to differentiate disjointed regions scattering at the same ($\theta, 2\theta$) condition as can be seen in Fig.~\ref{fig4}c. As ($\theta, 2\theta$) is varied, we see that there is a continuity in the regions coming in and out of the Bragg condition. The systematic comparison of the imaged 3D structures at different ($\theta, 2\theta$) values and their consistent evolution with respect to each other provides further confidence in the ability of our coded aperture reconstruction process in achieving accurate depth resolution.

\begin{figure*}[!t]
\centering
\includegraphics{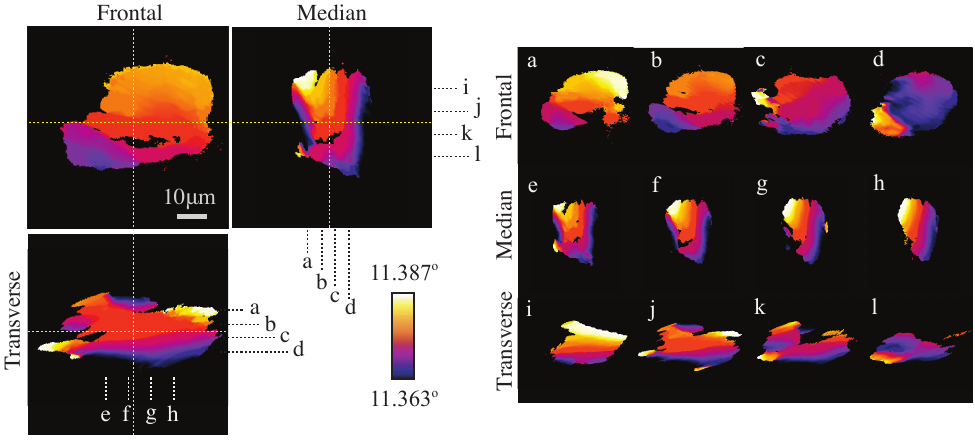}
\caption{\textbf{3D orientation reconstruction from rocking curve scanning.} The frontal, transverse, and median planes on the left match the cuts shown in Fig.~\ref{fig4} for consistent orientation across figures. To provide a more comprehensive analysis of the crystal structure, additional slices (a–l) have been introduced. These cross-sectional views of the 3D sample show smooth and continuous structural variations throughout the sample, providing a consistent depth reconstruction.}
\label{fig5}
\end{figure*}

The scans performed in Fig.~\ref{fig4} allow us to construct a 3D composite map of the local lattice tilt in this embedded domain. After reconstructing the 3D structure for each angle $\theta$ at a fixed $2\theta$, we calculated for each voxel the center of mass value of the $\theta$ angle. As seen in Fig.~\ref{fig4}c, as the tilt angle changes, different portions of the domain move into and out of Bragg conditions. The composite map of the center of mass of $\theta$ therefore covers the volume of the entire domain, leading to a larger imaged volume than a single tilt angle (see the difference in size of Fig.~\ref{fig4}a compared with Fig. \ref{fig5}). This procedure provides the dominant lattice tilt at that selected spatial position and allows the creation of a 3D orientation map for the sample. The corresponding reconstructed lattice orientation is shown in Fig.~\ref{fig5}. The slices indicated on the left panel reveal a spread of $0.02^{\circ}$ of the tilt in the internal structure of the embedded orthorhombic domain (Fig.~\ref{fig5}a-l). The shape of the domain and the gradients achieved from the composite center of mass images (Fig.~\ref{fig5}) agree well with the observed changes in the domain from Fig.~\ref{fig4}c. This continuous gradient of lattice tilts across the internal components of the domain provides further verification of this method. 

Visualizing the spatial distribution of lattice tilt within a domain in 3D yields important information about its internal structure. In addition, our 3D method can be combined with different types of scans to probe a variety of structural properties, such as using a longitudinal scan to visualize the axial strain and its gradients within a domain in 3D. In sum, our results demonstrate the power of our novel depth-sensitive imaging technique in uncovering deeply embedded structures on the mesoscale with nanoscale resolution.

\section*{Discussion}

In this work we have introduced a novel structured illumination approach to DFXM that enables 3D diffraction contrast imaging with a fixed sample and detector geometry, designed for the investigation of crystalline samples. Our innovative technique replaces the traditional sample rotation or thin-beam slicing in existing DFXM methods with a scanning aperture featuring structured patterns. This advancement opens up exciting possibilities for the reconstruction of heterogeneous lattice structure within crystals with sub-micron 3D resolution. The potential of this method for revealing relevant local structure pertinent to crystalline materials in-situ was shown via 3D reconstruction of a twin domain in Cu-Ba122 at cryogenic temperatures.

Our approach can also benefit various other x-ray microscopy techniques. For example, Laue micro-diffraction microscopy has already demonstrated the effective use of micro-coded apertures to expedite data collection times \cite{gursoy2022depth, gursoy2023digital}. Likewise, diffraction imaging at grazing-incidence angles shares similarities, allowing for the manipulation of the diffracted beam through a coded aperture to either eliminate the need for rotations or reduce their frequency. Furthermore, x-ray fluorescence microscopy could improve with a coded aperture scanning approach in conjunction with a pixelated energy-resolving detector, which would negate the need for rotation altogether. However, it is worth noting that while this approach shows promise for these techniques, each of them does come with its specific experimental constraints that need to be considered and addressed.

For a widespread application of our approach as alluded to above we discuss potential future directions of handle the optimization problem outlined in Eq.~\ref{eq3}. Although we employed a least-squares method, a better alternative approach could involve utilizing a regularized version of the problem \cite{gursoy2015maximum}. Introducing a cost function that penalizes infeasible or unlikely solutions can offer a more effective foundation for compressive sensing \cite{donoho2006compressed, candes2006robust}. This could potentially result in accelerated data acquisition with a reduced requirement for capturing a large number of images. Depending on the availability of training data, we can also explore the potential benefits of employing supervised \cite{yang2018low, liu2020tomogan} or unsupervised \cite{ulyanov2018deep} models based on deep neural networks. These models have the capacity to enhance both accuracy and reduce the number of required measurements and radiation dose, depending on the specific application. The signal reconstruction problem, implemented here, is addressed individually for each pixel intensity modulation profile resulting from the scanning of the coded aperture. As a result, the computational complexity of this method scales linearly with the quantity of detector pixels, making it amenable to parallel processing. This approach offers a direct avenue for harnessing the computational power of GPUs, which offer thousands of parallel processing units in contrast to CPUs. In terms of memory requirements, the method also scales linearly with the number of detector pixels; however, this demand can be partly facilitated by caching memory that are shared by all processors. 
 
In this study, we introduced a coded aperture approach for stable 3D imaging without sample rotation, providing detailed reconstructions while maintaining sample stability, which is particularly beneficial for complex or alignment-sensitive samples. This method enables full-field 3D views of structural properties, offering new insights into bulk phenomena such as charge density waves and magnetic phase competition, while also supporting rapid, low-dose data collection. Although optimizing aperture effects or resolution was beyond this study’s scope, future research could refine these aspects to enhance its utility alongside existing diffraction-based techniques. Furthermore, while flux reduction is a current trade-off, future work could explore non-attenuating coded apertures that leverage the flux and lift the resolution limitations imposed by flux reduction through spatial encoding as phase shifts in a coherent beam. This approach holds strong potential for expanding the capabilities of diffraction-based techniques in materials research and advancing coherent imaging systems.

In this study, we introduced a coded aperture approach for stable 3D imaging without sample rotation, achieving detailed reconstructions while maintaining sample stability, which is particularly advantageous for complex or alignment-sensitive samples. This method enables full-field 3D views of structural properties, providing new insights into bulk phenomena such as charge density waves and magnetic phase competition, while also supporting rapid, low-dose data collection. Although optimizing the aperture’s effects or resolution was beyond this study’s scope, future research could focus on refining these aspects to increase its utility alongside existing diffraction-based techniques.

Looking forward, our approach and ongoing efforts include the development of non-attenuating or mixed coded apertures that increase photon efficiency and support advanced encoding schemes through complex wave interactions. By utilizing both phase shifts and intensity modulation, these apertures allow for more refined manipulation of Bragg-diffracted signals, overcoming flux reduction limitations. This direction holds significant potential to expand DFXM in materials research, increasing its utility alongside existing diffraction-based methods to capture high-resolution 3D structural details and dynamic material behaviors.

\section*{Methods}

\subsection*{Sample Preparation}

We grew the Cu-Ba122 crystal used in this work using a self-flux method \cite{chu2009determination, ni2010temperature}. The crystal grows as thick platelets, with the large face corresponding to the tetragonal $ab$-plane. The \ce{Cu} doping level $0.02 \pm 0.0008$ of the batch was determined by electron probe micronanalysis wavelength-dispersive spectroscopy (EPMA-WDS) using parent compound BaFe$_{2}$As$_2$ and Cu metal as calibration samples. We obtained the measured sample by cleaving it from a single crystal using a razor blade, and it had the dimensions \SI{3.9}{\milli\meter} $\times$ \SI{1.4}{\milli\meter} $\times$ \SI{85}{\micro\meter}.

\subsection*{Dark-Field X-ray Microscope}

At the 6-ID-C beamline of the APS, DFXM employed a CRL made of beryllium \cite{qiao2020large} as the focusing optics, achieving illumination of the volume of interest with a spot size of \SI{50}{\micro\meter} on the sample. The setup incorporated a polymeric CRL \cite{nazmov2004fabrication} as an objective lens with a \SI{131}{\milli\meter} focal length at \SI{20}{\kilo\electronvolt}. The sample was placed \SI{140}{\milli\meter} from the objective lens, experiencing an x-ray magnification of $24$ incident upon a LuAG(Ce) scintillator crystal with a thickness of \SI{20}{\micro\meter}. An optical imaging system was used to record the scintillator surface and provided an additional magnification of $5$, leading to an overall magnification of $120$. Consequently, we achieved a pixel size of \SI{50}{\nano\meter} in the detector plane. 

\subsection*{Coded Aperture Design and Manufacturing}

The pattern for the coded aperture was derived from a de Bruijn sequence of length 256 and order 8, providing uniqueness within each 8-bit long sequence. The fabrication of the coded aperture took place in the clean room at the Center for Nanoscale Materials at Argonne National Laboratory, utilizing direct-write lithography and electroplating. It featured a \SI{5}{\micro\meter} wide \ce{Au} for the 1-bits and \SI{10}{\micro\meter} wide silicon nitride (\ce{Si3N4}) membrane with negligible absorption for the 0-bits. Thickness measurements along the optical axis indicated an approximate thickness of \SI{5}{\micro\meter} using a scanning electron microscope and scanning profilometer.

\subsection*{Data Collection}

We positioned the coded aperture \SI{9.4}{\milli\meter} upstream of the sample, as close to the sample as could be achieved before contacting the outer vacuum shroud of our cryostat. This configuration ensures that the barcode footprint of the coded aperture projects onto the sample. The aperture was scanned perpendicular to the incident beam direction within the horizontal scattering plane in increments of \SI{1}{\micro\meter} steps for a total scan range of \SI{225}{\micro\meter}. The translation stage used demonstrated positional stability down to \SI{30}{\nano\meter}. The motor encoder accurately tracked these positions, which were utilized in the subsequent image reconstruction. The exposure of the detector was \SI{0.5}{\milli\second} at each coded aperture step scan location. 

We collected a matrix of coded aperture scanning measurements at ten different sample incident angles from \SI{-0.014}{\degree} to \SI{0.013}{\degree} in steps of \SI{0.003}{\degree} from $(040)_o$ Bragg peak condition, and three different scattering angles, \SI{-0.003}{\degree}, \SI{0}{\degree} and \SI{0.003}{\degree} from Bragg peak $2\theta$ of this twin domain. This measurement generated thirty datasets, with each dataset comprising 225 images.

\subsection*{Sample Reconstruction}

We performed digital sample reconstruction from the acquired data using custom in-house software. This software solves the optimization problem described in Eq.~\ref{eq3} and employs the non-negative least squares method \cite{lawson1995solving}, implemented in SciPy \cite{virtanen2020scipy}. The effective resolution is calculated to be \SI{0.45}{\micro\meter} for the raw projection images and \SI{0.49}{\micro\meter} for the 3D reconstruction along the scattering angle, based on the Fourier Ring Correlation with half-bit measurement, as implemented in ImageJ \cite{nieuwenhuizen2013measuring}. To construct the system matrix ($\mathbf{A}$), we utilized the actual positions of the coded aperture based on the encoder readings of the high-precision linear stage used for aperture translation. The reconstruction took several hours on a commercial laptop without parallelization or code optimization. Currently, we are finalizing the software repository but can provide the code, scripts, and data upon request.

\section*{Acknowledgements} This research used resources of the Advanced Photon Source and the Center for Nanoscale Materials, U.S. Department of Energy (DOE) Office of Science User Facilities and is based on work supported by Laboratory Directed Research and Development (LDRD) funding from Argonne National Laboratory, provided by the Director, Office of Science, of the U.S. DOE under Contract No. DE-AC02-06CH11357. The work of M.H. and S.H. was supported by the U.S. Department of Energy (DOE), Office of Science, Basic Energy Sciences (BES), Materials Science and Engineering Division. The work of K.A.Y. and I.R.F. on crystal growth and characterization received support from the DOE, Office of Science, BES, under contract DE-AC02-76SF00515. The authors thank the Karlsruhe Nano Micro Facility (KNMF) for the fabrication of the polymer x-ray optics. 

\bibliographystyle{ieeetr}
\bibliography{refs}

\begin{thebibliography}{10}

\bibitem{simons2015dark}
H.~Simons, A.~King, W.~Ludwig, C.~Detlefs, W.~Pantleon, S.~Schmidt,
  F.~St{\"o}hr, I.~Snigireva, A.~Snigirev, and H.~F. Poulsen, ``Dark-field
  x-ray microscopy for multiscale structural characterization,'' {\em Nature
  Communications}, vol.~6, no.~1, p.~6098, 2015.

\bibitem{pfeiffer2008hard}
F.~Pfeiffer, M.~Bech, O.~Bunk, P.~Kraft, E.~F. Eikenberry, C.~Br{\"o}nnimann,
  C.~Gr{\"u}nzweig, and C.~David, ``Hard-x-ray dark-field imaging using a
  grating interferometer,'' {\em Nature materials}, vol.~7, no.~2,
  pp.~134--137, 2008.

\bibitem{wang2015hard}
H.~Wang, Y.~Kashyap, and K.~Sawhney, ``Hard-x-ray directional dark-field
  imaging using the speckle scanning technique,'' {\em Physical Review
  Letters}, vol.~114, no.~10, p.~103901, 2015.

\bibitem{jakobsen2019mapping}
A.~Jakobsen, H.~Simons, W.~Ludwig, C.~Yildirim, H.~Leemreize, L.~Porz,
  C.~Detlefs, and H.~Poulsen, ``Mapping of individual dislocations with
  dark-field x-ray microscopy,'' {\em Journal of Applied Crystallography},
  vol.~52, no.~1, pp.~122--132, 2019.

\bibitem{davis1995phase}
T.~J. Davis, D.~Gao, T.~Gureyev, A.~Stevenson, and S.~Wilkins, ``Phase-contrast
  imaging of weakly absorbing materials using hard x-rays,'' {\em Nature},
  vol.~373, no.~6515, pp.~595--598, 1995.

\bibitem{yildirim2020probing}
C.~Yildirim, P.~Cook, C.~Detlefs, H.~Simons, and H.~F. Poulsen, ``Probing
  nanoscale structure and strain by dark-field x-ray microscopy,'' {\em MRS
  Bulletin}, vol.~45, no.~4, pp.~277--282, 2020.

\bibitem{poulsen2017x}
H.~F. Poulsen, A.~Jakobsen, H.~Simons, S.~R. Ahl, P.~Cook, and C.~Detlefs,
  ``X-ray diffraction microscopy based on refractive optics,'' {\em Journal of
  Applied Crystallography}, vol.~50, no.~5, pp.~1441--1456, 2017.

\bibitem{breckling2022automated}
S.~Breckling, B.~Kozioziemski, L.~Dresselhaus-Marais, A.~Gonzalez, A.~Williams,
  H.~Simons, P.~Chow, and M.~Howard, ``An automated approach to the alignment
  of compound refractive lenses,'' {\em Synchrotron Radiation}, vol.~29, no.~4,
  pp.~947--956, 2022.

\bibitem{gursoy2017rapid}
D.~G{\"u}rsoy, Y.~P. Hong, K.~He, K.~Hujsak, S.~Yoo, S.~Chen, Y.~Li, M.~Ge,
  L.~M. Miller, Y.~S. Chu, {\em et~al.}, ``Rapid alignment of nanotomography
  data using joint iterative reconstruction and reprojection,'' {\em Scientific
  Reports}, vol.~7, no.~1, p.~11818, 2017.

\bibitem{forbes2021structured}
A.~Forbes, M.~de~Oliveira, and M.~R. Dennis, ``Structured light,'' {\em Nature
  Photonics}, vol.~15, no.~4, pp.~253--262, 2021.

\bibitem{shapiro2008computational}
J.~H. Shapiro, ``Computational ghost imaging,'' {\em Physical Review A},
  vol.~78, no.~6, p.~061802, 2008.

\bibitem{duarte2008single}
M.~F. Duarte, M.~A. Davenport, D.~Takhar, J.~N. Laska, T.~Sun, K.~F. Kelly, and
  R.~G. Baraniuk, ``Single-pixel imaging via compressive sampling,'' {\em IEEE
  Signal Processing Magazine}, vol.~25, no.~2, pp.~83--91, 2008.

\bibitem{snigirev1996compound}
A.~Snigirev, V.~Kohn, I.~Snigireva, and B.~Lengeler, ``A compound refractive
  lens for focusing high-energy x-rays,'' {\em Nature}, vol.~384, no.~6604,
  pp.~49--51, 1996.

\bibitem{gorelick2011high}
S.~Gorelick, J.~Vila-Comamala, V.~A. Guzenko, R.~Barrett, M.~Salom{\'e}, and
  C.~David, ``High-efficiency fresnel zone plates for hard x-rays by 100 kev
  e-beam lithography and electroplating,'' {\em Journal of Synchrotron
  Radiation}, vol.~18, no.~3, pp.~442--446, 2011.

\bibitem{paglione2010high}
J.~Paglione and R.~L. Greene, ``High-temperature superconductivity in
  iron-based materials,'' {\em Nature physics}, vol.~6, no.~9, pp.~645--658,
  2010.

\bibitem{fernandes2022iron}
R.~M. Fernandes, A.~I. Coldea, H.~Ding, I.~R. Fisher, P.~Hirschfeld, and
  G.~Kotliar, ``Iron pnictides and chalcogenides: a new paradigm for
  superconductivity,'' {\em Nature}, vol.~601, no.~7891, pp.~35--44, 2022.

\bibitem{rotter2008superconductivity}
M.~Rotter, M.~Tegel, and D.~Johrendt, ``Superconductivity at 38 k in the iron
  arsenide (ba 1- x k x) fe 2 as 2,'' {\em Physical Review Letters}, vol.~101,
  no.~10, p.~107006, 2008.

\bibitem{tanatar2009direct}
M.~Tanatar, A.~Kreyssig, S.~Nandi, N.~Ni, S.~L. Bud’ko, P.~Canfield,
  A.~Goldman, and R.~Prozorov, ``Direct imaging of the structural domains in
  the iron pnictides a fe 2 as 2 (a= ca, sr, ba),'' {\em Physical Review B},
  vol.~79, no.~18, p.~180508, 2009.

\bibitem{blomberg2012effect}
E.~Blomberg, A.~Kreyssig, M.~A. Tanatar, R.~Fernandes, M.~Kim, A.~Thaler,
  J.~Schmalian, S.~Bud'Ko, P.~Canfield, A.~Goldman, {\em et~al.}, ``Effect of
  tensile stress on the in-plane resistivity anisotropy in bafe 2 as 2,'' {\em
  Physical Review B}, vol.~85, no.~14, p.~144509, 2012.

\bibitem{plumb2023dark}
J.~Plumb, I.~Poudyal, R.~L. Dally, S.~Daly, S.~D. Wilson, and Z.~Islam, ``Dark
  field x-ray microscopy below liquid-helium temperature: The case of namno2,''
  {\em Materials Characterization}, vol.~204, p.~113174, 2023.

\bibitem{gursoy2022depth}
D.~G{\"u}rsoy, D.~Sheyfer, M.~Wojcik, W.~Liu, and J.~Z. Tischler,
  ``Depth-resolved {Laue} microdiffraction with coded apertures,'' {\em Journal
  of Applied Crystallography}, vol.~55, no.~5, 2022.

\bibitem{gursoy2023digital}
D.~G{\"u}rsoy, D.~Sheyfer, M.~Wojcik, W.~Liu, and J.~Z. Tischler, ``Digital
  autofocusing of a coded-aperture {Laue} diffraction microscope,'' {\em Review
  of Scientific Instruments}, vol.~94, no.~1, 2023.

\bibitem{gursoy2015maximum}
D.~G{\"u}rsoy, T.~Bi{\c{c}}er, J.~D. Almer, R.~Kettimuthu, S.~R. Stock, and
  F.~De~Carlo, ``Maximum a posteriori estimation of crystallographic phases in
  x-ray diffraction tomography,'' {\em Philosophical Transactions of the Royal
  Society A: Mathematical, Physical and Engineering Sciences}, vol.~373,
  no.~2043, p.~20140392, 2015.

\bibitem{donoho2006compressed}
D.~L. Donoho, ``Compressed sensing,'' {\em IEEE Transactions on Information
  Theory}, vol.~52, no.~4, pp.~1289--1306, 2006.

\bibitem{candes2006robust}
E.~J. Cand{\`e}s, J.~Romberg, and T.~Tao, ``Robust uncertainty principles:
  Exact signal reconstruction from highly incomplete frequency information,''
  {\em IEEE Transactions on Information Theory}, vol.~52, no.~2, pp.~489--509,
  2006.

\bibitem{yang2018low}
X.~Yang, V.~De~Andrade, W.~Scullin, E.~L. Dyer, N.~Kasthuri, F.~De~Carlo, and
  D.~G{\"u}rsoy, ``Low-dose x-ray tomography through a deep convolutional
  neural network,'' {\em Scientific Reports}, vol.~8, no.~1, p.~2575, 2018.

\bibitem{liu2020tomogan}
Z.~Liu, T.~Bicer, R.~Kettimuthu, D.~Gursoy, F.~De~Carlo, and I.~Foster,
  ``Tomogan: low-dose synchrotron x-ray tomography with generative adversarial
  networks: discussion,'' {\em JOSA A}, vol.~37, no.~3, pp.~422--434, 2020.

\bibitem{ulyanov2018deep}
D.~Ulyanov, A.~Vedaldi, and V.~Lempitsky, ``Deep image prior,'' in {\em
  Proceedings of the IEEE Conference on Computer Vision and Pattern
  Recognition}, pp.~9446--9454, 2018.

\bibitem{chu2009determination}
J.-H. Chu, J.~G. Analytis, C.~Kucharczyk, and I.~R. Fisher, ``Determination of
  the phase diagram of the electron-doped superconductor ba (fe 1- x co x) 2 as
  2,'' {\em Physical Review B}, vol.~79, no.~1, p.~014506, 2009.

\bibitem{ni2010temperature}
N.~Ni, A.~Thaler, J.~Yan, A.~Kracher, E.~Colombier, S.~Bud’Ko, P.~Canfield,
  and S.~Hannahs, ``Temperature versus doping phase diagrams for ba (fe 1- x tm
  x) 2 as 2 (tm= ni, cu, cu/co) single crystals,'' {\em Physical Review B},
  vol.~82, no.~2, p.~024519, 2010.

\bibitem{qiao2020large}
Z.~Qiao, X.~Shi, P.~Kenesei, A.~Last, L.~Assoufid, and Z.~Islam, ``A large
  field-of-view high-resolution hard x-ray microscope using polymer optics,''
  {\em Review of Scientific Instruments}, vol.~91, no.~11, 2020.

\bibitem{nazmov2004fabrication}
V.~Nazmov, E.~Reznikova, J.~Mohr, A.~Snigirev, I.~Snigireva, S.~Achenbach, and
  V.~Saile, ``Fabrication and preliminary testing of x-ray lenses in thick su-8
  resist layers,'' {\em Microsystem Technologies}, vol.~10, pp.~716--721, 2004.

\bibitem{lawson1995solving}
C.~L. Lawson and R.~J. Hanson, {\em Solving least squares problems}.
\newblock SIAM, 1995.

\bibitem{virtanen2020scipy}
P.~Virtanen, R.~Gommers, T.~E. Oliphant, M.~Haberland, T.~Reddy, D.~Cournapeau,
  E.~Burovski, P.~Peterson, W.~Weckesser, J.~Bright, {\em et~al.}, ``{SciPy}
  1.0: fundamental algorithms for scientific computing in python,'' {\em Nature
  methods}, vol.~17, no.~3, pp.~261--272, 2020.

\bibitem{nieuwenhuizen2013measuring}
R.~P. Nieuwenhuizen, K.~A. Lidke, M.~Bates, D.~L. Puig, D.~Gr{\"u}nwald,
  S.~Stallinga, and B.~Rieger, ``Measuring image resolution in optical
  nanoscopy,'' {\em Nature methods}, vol.~10, no.~6, pp.~557--562, 2013.

\end{thebibliography}

\end{document}